\documentclass[12pt]{article}
\usepackage{a4wide,amssymb,cite}
\usepackage{a4wide,amssymb,cite,graphicx}
\usepackage{epsfig}
\usepackage[usenames,dvipsnames]{color}
\usepackage{slashed}

\newcommand{\be}{\begin{equation}}
\newcommand{\ee}{\end{equation}}
\newcommand{\bea}{\begin{eqnarray}}
\newcommand{\eea}{\end{eqnarray}}

\def\circa#1{\,\raise.3ex\hbox{$#1$\kern-.75em\lower1ex\hbox{$\sim$}}\,}

\begin{document}

\begin{titlepage}
%
%


%

\begin{centering}
\vspace{1cm}
{\large {\bf Chaotic inflation and unitarity problem}} \\

\vspace{1.5cm}

 {\bf Hyun Min Lee} \\
\vspace{.5cm}

{\it Department of Physics, Chung-Ang University, Seoul 156-756, Korea}  \\

\vspace{.1in}

\end{centering}
\vspace{2cm}

\begin{abstract}
\noindent
We consider a general chaotic inflation model with non-canonical kinetic term, resulting in attractor solutions for the inflation of quadratic or other monomial type. 
In particular, the form of the kinetic term and the potential is fixed due to the requirement that the inflation model is a quadratic form in the large field values of the inflaton. We show that a large coupling in the non-canonical kinetic term allows for the slow-roll inflation with sub-Planckian field values of the inflaton and the successful predictions of the quadratic or other monomial type chaotic inflation in light of BICEP2 results are maintained in our model.
We find that due to the large rescaling of the inflaton field in the vacuum, there is no unitarity problem below the Planck scale. 

\end{abstract}

\vspace{3cm}

\end{titlepage}

\section{Introduction}

The BICEP2 collaboration \cite{bicep} has recently announced the evidence for B-modes in the CMB polarization, which are presumably originated from the primordial gravitational waves of cosmic inflation.
The reported value of the tensor-to-scalar ratio is $r=0.20^{+0.07}_{-0.05}$, which is quite larger than the previous upper bound, $r<0.11$,  given at $95\%$ C.L. by Planck data combined with WMAP polarization and high-$l$ \cite{planck}.  In the case of single-field inflation models, such a large value of $r$ implies that there was an excursion of the inflaton  beyond the Planck-scale field values during inflation \cite{lyth}.  As a result, simple monomial chaotic inflation models have become favored \footnote{We note that there is an issue on a sizable running of the spectral index, which resolves the tension between BICEP2 and Planck \cite{bicep}, but it is hard to realize it in most of slow-roll inflation models. }.
On the other hand, many of the inflation models predicting a small value of $r$ including Starobinsky model \cite{starobinsky} and Higgs inflation \cite{higgsinf1} in a simple form \footnote{See, however, the recent discussion in Ref.~\cite{higgsinf2,higgsinf3,takahashi}.} have been disfavored by a large $r$, on top of the problem of unitarity violation \cite{unitarity,trott,riotto,gian}.

Among the monomial chaotic inflation models, the quadratic inflation \cite{linde} has drawn a new attention due to the fact that the model predictions are consistent with Planck$+$ BICEP2 within $1\sigma$.
We consider a general quadratic inflation with a single scalar field containing the general kinetic term and potential. The condition that the quadratic inflation is reproduced at large field values fixes the form of the kinetic and potential terms. As a consequence, we show how the model parameters in the general quadratic inflation are constrained after BICEP2. 

A large $r$ suggests that the inflaton field with canonical kinetic term must have travelled to trans-Planckian field values during inflation, so there is a concern about how to address the quantum gravity effects suppressed by the Planck scale from the pointview of the effective field theory. In our general quadratic inflation, the field value of the non-canonical inflaton remains sub-Planckian during inflation, thanks to a large coupling in the non-canonical kinetic term \cite{runkinetic,takahashi}. Furthermore, we find that there is no unitarity violation coming from the large coupling below the Planck scale, because the wave function rescaling of the inflaton field in the vacuum eliminates any positive powers of the large coupling as in the induced inflation models \cite{gian}. Therefore, the higher order terms for the non-canonical inflaton, that are suppressed by the Planck scale, can be safely ignored. 
We also generalize the results to the case with general monomial chaotic inflation in the model with a similar conclusion.
There were  previous discussions on more general polynomial chaotic inflation models obtaining a large tensor-to-scalar ratio with sub-Planckian inflation field values \cite{subPlanck}.

The paper is organized as follows.
We begin with a model description of the general chaotic inflation focusing on the quadratic form and discuss the inflation constraints on the model in view of Planck and BICEP2. Then, we address the issue of unitarity violation in this model and conclusions are drawn.

\section{General chaotic inflation}

We introduce a real scalar inflaton having a general kinetic term with two derivatives and a potential as follows,
\bea
\frac{{\cal L}}{\sqrt{-g}}=\frac{1}{2}\Omega(\phi)R- \frac{1}{2}K(\phi)(\partial_\mu\phi)^2 - V(\phi). \label{lag1}
\eea
Without loss of generality, we choose the general kinetic term and the potential in the following discussion as
\bea
\Omega(\phi)=1+\zeta h(\phi), \quad K(\phi)= 1+\xi f(\phi),\quad V(\phi)= \lambda (g(\phi)- g(\phi_0))^2 \label{functions}
\eea
where $\zeta,\xi,\lambda$ are constant parameters, $h(\phi), f(\phi), g(\phi)$ are general functions of  $\phi$, and we have chosen the potential to vanish in the vacuum with $\langle\phi\rangle=\phi_0$.
The generic feature of the above Lagrangian is that the inflaton field has non-canonical kinetic terms for $h(\phi_0)\neq 0$ or $f(\phi_0)\neq 0$ so the interaction terms get rescaled after the graviton or inflaton kinetic term is made canonical.
This aspect leads to an interesting result for the unitarity scale as will be discussed in a later section.
Here, we have used the units with $M_P=1$ but the Planck scale is introduced whenever necessary.

We note that the non-canonical kinetic term $K(\phi)$ for the inflaton could be always canonically normalized locally, as far as the inflaton field distance during inflation is not larger than the curvature radius in the field space of graviton and inflaton \cite{burgess}.  Then, one can describe a single-field inflation in terms of the canonically normalized field.  As a consequence, when the canonically normalized field describes monomial chaotic inflation, it appears super-Planckian effectively, even if the original inflaton field before a canonical normalization is sub-Planckian.  The case with a large non-minimal coupling has been discussed in the context of Starobinsky-like models in Ref.~\cite{gian}, but the tensor-to-scalar ratio in this case turned out to be negligible. Thus, for discussion on the scalar functions at the leading order and correction terms in this work, we reside in the field basis where the non-minimal coupling of order unity is present and the inflaton kinetic term is non-canonical.
Then, we focus on the case that the non-canonical kinetic term $K(\phi)$ plays a crucial role during inflation while the non-minimal coupling does not.

\subsection{Quadratic inflation}

Even if the potential is not dominated by a mass term, the model can describe a quadratic inflation due to a non-canonical kinetic term during inflation. For simplicity, we take the functions $f(\phi)$, $g(\phi)$ and $h(\phi)$ to be power-like,
\bea
f(\phi)=\phi^n,\quad g(\phi)= \phi^m,  \quad h(\phi)= \phi^k .\label{powers}
\eea  
Then, in the limit of $\xi f(\phi)=\xi \phi^n \gg 1$ during inflation \footnote{In the opposite case with $\xi f(\phi)\ll 1$, our model is equivalent to chaotic inflation models with canonical kinetic term.}, we obtain the canonically normalized inflaton as 
$\chi\approx \sqrt{\xi} \phi^{n/2+1}/(n/2+1)$, as far as the mixing between graviton and inflaton can be ignored for $\zeta \phi^k \ll 1$, namely, $\zeta/\xi \ll \phi^{n-k} $ for $\zeta={\cal O}(1)$ and a large $\xi$.
 In this case, the potential function becomes $g(\phi(\chi))\sim \chi^{2m/(n+2)}$.  Therefore, choosing $m=(n+2)/2$ and ignoring the contribution of the non-minimal coupling, we  obtain the Lagrangian during inflation as follows,
\bea
\frac{{\cal L}}{\sqrt{-g}}\approx \frac{1}{2}R-\frac{1}{2}(\partial_\mu\chi)^2 - \frac{\lambda}{\xi} \Big(1+\frac{n}{2}\Big)^2 (\chi- \chi_0)^2
\eea
where $\chi_0\equiv \sqrt{\xi}\phi^{1+n/2}_0/(1+n/2)$.
Then, we can get rid of $\chi_0$ by making a shift for $\chi$ without changing any physics, ending up with a quadratic inflation model. The running kinetic term with $f(\phi)=\phi^2$ and a general form was previously considered in Ref.~\cite{runkinetic,takahashi} but the unitarity issue was not discussed. Unitarity problem distinguishes the models as will be shown in a later section. 

The mass of the canonical inflaton is given by 
\be
m_\chi=\frac{n+2}{2} \sqrt{\frac{2\lambda}{\xi}}.  \label{inflatonmass}
\ee
Therefore, the predictions of the model for inflation are the same as for the usual quadratic inflation with canonical kinetic term. Although variations of quadratic inflation with a polynomial potential can be also consistent with Planck $+$ BICEP2 \cite{seto}, we focus on the simplest case with attractor solutions for the quadratic inflation. 

The slow-roll parameters for the quadratic inflation at horizon exit are written in terms of the number of efoldings $N$ as
\bea
\epsilon_*= \eta_*=\frac{2}{\chi^2_*}=\frac{1}{2N+1}.
\eea
Then, we get the spectral index of the scalar perturbation and the tensor-to-scalar ratio, respectively, as 
\bea
n_s&=&1-6\epsilon_*+2\eta_*=1-\frac{4}{2N+1},\\
r&=&16\epsilon_* = \frac{16}{2N+1}.
\eea
Therefore, for $N=50(60)$, we find that $n_s=0.960(0.967)$ and $r=0.158(0.132)$.
As a result, the model leads to large primordial gravitational waves, and the model predictions can be consistent with $n_s=0.9600\pm 0.0071$ and $r=0.20^{+0.07}_{-0.05}$ given by Planck$+$WP$+$high-$l+$BICEP2 within about $1\sigma$.
On the other hand, the running of the spectral index is negligibly small as 
\bea
\frac{d n_s}{d\ln k} = -7.84(5.46)\times 10^{-4}.
\eea
Thus, there is a tension with the previous limit on the tensor-to-scalar ratio, $r<0.11$, obtained at $95\%$ C. L. for Planck$+$WP+high-$l$, in the case of a zero running of the spectral index.
But, we do not discuss the solution to resolve the tension in this work.

Finally, from the COBE normalization, $A_s=V/(24\pi^2 M^4_P \epsilon_*)=2.196\times 10^{-9}$, 
the inflaton mass is determined 
\be
m_\chi=1.74(1.45)\times 10^{13}\,{\rm GeV},
\ee
for $N=50(60)$,
resulting in the fixed ratio of the constant parameters as
\bea
\frac{\lambda}{\xi}=\frac{16}{(n+2)^2}\times 6.37(4.44)\times 10^{-12}.  \label{cobe1}
\eea
As a consequence, we need a large $\xi$ or a small $\lambda$ to satisfy the COBE normalization.
For a higher power of the polynomial, we would get a smaller ratio $\lambda/\xi$.
The value of the inflaton mass is suggestive of solving the vacuum instability problem with a heavy scalar threshold in the SM with an inflaton coupling to the Higgs doublet \cite{vsb}.

In particular, for $f(\phi)=\phi^2$ and $g(\phi)=\phi^2$, the Lagrangian (\ref{lag1}) with eq.(\ref{functions}) becomes 
\be
\frac{{\cal L}}{\sqrt{-g}}= \frac{1}{2}R-\frac{1}{2}(1+\xi\phi^2)(\partial_\mu\phi)^2 -\lambda(\phi^2-\phi^2_0)^2.
\ee
In this case, we can get the analytic expression for the canonical inflation field as follows,
\bea
\chi= \frac{1}{2\sqrt{\xi}} \ln\left(\frac{\sqrt{\sqrt{1+\xi\phi^2}+1}+\sqrt{\sqrt{1+\xi\phi^2}-1}}{\sqrt{\sqrt{1+\xi\phi^2}+1}-\sqrt{\sqrt{1+\xi\phi^2}-1}}\right)+\frac{1}{2} \phi \sqrt{1+\xi\phi^2}.
\eea
For $\xi \phi^2\gg 1$ during inflation, we obtain $\chi\approx \sqrt{\xi}\phi^2/2$.
The slow-roll condition with $\chi\gg 1$, however, requires a stronger condition on the $\phi$ field value as $\phi\gg 1/\xi^{1/4}$.  
Therefore, for a large $\xi$, there is a large room, $1/\xi^{1/4}\ll \phi \ll 1$, for the $\phi$ field value to be sub-Planckian during inflation.

In passing, we make some comments on the higher order Planck-suppressed terms in the effective field theory. Keeping the $Z_2$ symmetry with $\phi\rightarrow -\phi$ and the second order derivatives,  we can consider higher order terms for the original field $\phi$ or the redefined field $\chi$, respectively, as follows,
\bea
\phi:\quad&& \sum_n \left( a_n\,\frac{\phi^{2+2n}}{M^{2+2n}_P}(\partial\phi)^2+ b_n \,\frac{\phi^{4+2n}}{M^{2n}_P} \right),   \label{hdo1}\\
\chi:\quad&& \sum_n \left( c_n\, \frac{\chi^n}{M^n_P}(\partial\chi)^2+ d_n \,\frac{\chi^{n+2}}{M^{n-2}_P}\right). \label{hdo2}
\eea
Here, we assumed that the unitarity cutoff is given by the Planck scale.
Then,  using $\chi\approx \sqrt{\xi}\phi^2/2$ for $\xi \phi^2\gg 1$ during inflation, the higher order terms for the $\chi$ field in eq.~(\ref{hdo2}) can be rewritten as
\be
\chi:\quad \sum_n\left(c_n\,\frac{\xi^{1+\frac{n}{2}}}{2^n}\frac{\phi^{2+2n}}{M^{2+2n}_P} (\partial\phi)^2 + \frac{d_n \xi^{1+\frac{n}{2}}}{2^{n+2}}\,\frac{\phi^{4+2n}}{M^{2n}_P}  \right).  \label{hd022}
\ee
Therefore, in order for the higher order terms to be frame independent, we need the following conditions on the coefficients,
\be
a_n\simeq \frac{c_n\xi^{1+\frac{n}{2}}}{2^n},\quad\quad  b_n\simeq \frac{d_n \xi^{1+\frac{n}{2}}}{2^{n+2}}.
\ee
Then, since the $\phi$ field is sub-Planckian, the higher order corrections suppressed by the Planck scale can be ignored during inflation, as far as $a_n={\cal O}(1)$ and $b_n={\cal O}(1)$. 
We note that if the vacuum expectation value of the inflaton is small,  the unitarity cutoff can be smaller than the Planck scale, even for the same non-minimal kinetic term in the leading order, as will be shown in a later section. Then, the higher order terms with a unitarity cutoff smaller than the Planck scale could be dangerous, because $\phi\gg \xi^{-1/4}$. 
We will come back to this issue with unitarity problem for general quadratic inflation models in a later section.

We note that for a linear form with $f(\phi)=\phi$, which is obviously the lowest term in the non-canonical kinetic term for a singlet scalar $\phi$, a quadratic inflation can be obtained only for a fractional power term in the potential as $g(\phi)=\phi^{3/2}$.

For the general polynomials satisfying $f(\phi)=\phi^n$ and $g(\phi)=\phi^{(n+2)/2}$, the potential for the canonically-normalized inflaton becomes a quadratic form, because
the canonical scalar field during inflation is written as $\chi\approx \sqrt{\xi} \phi^{n/2+1}/(n/2+1)$.
Then, the slow-roll condition, $\chi \gg 1$, again requires the $\phi$ field value to be $\phi\gg \xi^{-1/(n+2)}$.
As a consequence, for $\xi\sim 10^{l}$ and taking $\xi^{-1/(n+2)}\lesssim 10^{-2}$ for sub-Planckian $\phi$ field values, the power of the polynomial is restricted to $n\lesssim l/2-2$, or $\xi\gtrsim 10^{2(n+2)}$ is required for a given power of the polynomial.

\subsection{Monomial chaotic inflation}

In this section, for completeness, we consider the monomial chaotic inflation other than quadratic inflation in our model and compare the model predictions to BICEP2.  As noted before, for an arbitrary choice of powers, $n$ and $m$, in eq.~(\ref{powers}), the power of the potential for the canonically-normalized inflaton takes a general form of chaotic inflation  as $V(\phi(\chi))\sim \chi^{4m/(n+2)}$ for large filed values as follows,
\bea
\frac{{\cal L}}{\sqrt{-g}}\approx \frac{1}{2}R-\frac{1}{2}(\partial_\mu\chi)^2 - \frac{\lambda}{\xi^{\frac{2m}{n+2}}} \Big(1+\frac{n}{2}\Big)^{\frac{4m}{n+2}} \Big(\chi^{\frac{2m}{n+2}}- \chi^{\frac{2m}{n+2}}_0\Big)^2.
\eea

For the effective potential, $V(\chi)=\alpha \Big(\chi^{k/2}-\chi^{k/2}_0\Big)^2$,  with $k=\frac{4m}{n+2}$,  the slow-roll parameters at horizon exit are
\bea
\epsilon_*&=&\frac{k^2}{2}\frac{1}{\chi^2_* \Big(1-(\chi_0/\chi_*)^{k/2}\Big)^2}, \\
\eta_*&=& \frac{k(k-1)-k(k/2-1)(\chi_0/\chi_*)^{k/2}}{\chi^2_*\Big(1-(\chi_0/\chi_*)^{k/2}\Big)^2}.
\eea
On the other hand, the number of efoldings is 
\be
N=\frac{1}{2k}\left[\chi^2_*\left(1-\frac{1}{1-k/4}\left(\frac{\chi_0}{\chi_*}\right)^{k/2} \right) -\chi^2_e\left(1-\frac{1}{1-k/4}\left(\frac{\chi_0}{\chi_e}\right)^{k/2} \right)  \right]
\ee
where $\chi_e$ is the value of the inflaton field at the end of inflation, given by the solution to $\chi_e \Big(1-(\chi_0/\chi_e)^{k/2}\Big)=k/\sqrt{2}$.
Then, for $\chi_0\ll \chi_e <\chi_*$, we get the slow-roll parameters in terms of the number of efoldings as 
\be
\epsilon_*\approx \frac{k^2}{2\chi^2_*} \approx \frac{k}{2(2N+k/2)},\quad \eta_*\approx \frac{k(k-1)}{\chi^2_*}\approx \frac{k-1}{2N+k/2}.
\ee
For a slow-roll inflation with general monomial terms, we need $\chi^2_*\approx k(2N+k/2)$ so the fact that $\chi_*\gg 1$ implies that the non-canonical inflaton field values are sub-Planckian during inflation only for $\xi^{-1/(n+2)}\ll \phi \ll 1$, similarly to the general quadratic inflation. 
Therefore, we can keep the non-canonical inflaton field sufficiently small  due to a large coupling $\xi$ and accommodate various monomial chaotic inflation models containing the canonical kinetic term. 

Consequently, the spectral index of the scalar perturbation and the tensor-to-scalar ratio are 
\be
n_s\approx1-\frac{k+2}{2N+k/2},\quad r\approx \frac{8k}{2N+k/2}.
\ee
On the other hand, for $N=50$, the COBE normalization restricts the inflaton parameters as
\be
\frac{\lambda}{\xi^{k/2}}\approx \frac{2^{k+2}}{(n+2)^k} \frac{k^{1-k/2}}{(2N+k/2)^{-1+k/2}}\cdot\,6.37\times 10^{-12}.  \label{cobe2}
\ee
Here, we note that for $k<2$, the coupling $\xi$ in the non-canonical kinetic term gets smaller, as compared to the case with quadratic inflation in eq.~(\ref{cobe1}).

For instance, for $n=1$, i.e. a linear non-canonical kinetic term, we get $k=\frac{4}{3}m$.
In this case, for $N=50$, we find that $(n_s,r)=(0.967,0.106)$, $(0.954,0.211))$, $(0.941,0.314)$,  in the order of $k=4/3,8/3, 4$.
Next, for $n=2$, we get $k=m$ so the power of the effective potential is given by the square root of the original potential. For $n=3$, we get $k=\frac{4}{5} m$ and find for $N=50$ that $(n_s,r)=(0.972,0.06)$, $(0.964,0.127)$, $(0.957,0.190)$, $(0.949,0.252)$, in the order of $k=4/5,8/5,12/5,16/5$. 
Therefore, as compared to Planck and BICEP2, successful chaotic inflation models with fractional power can be obtained for the general polynomial functions $f(\phi)$ and $g(\phi)$ with regular powers. Just for quadratic inflation, the running of the spectral index is too small to be observed at the current level of precision.

\section{General chaotic inflation and unitarity scale}

It is remarkable that the unitarity scale depends on the background field values and the vacuum expectation value of the inflaton field plays an important role in determining the unitarity scale \cite{uvhiggs,gian}.
In this section, focusing on the general monomial functions leading to the quadratic inflation and general monomial chaotic inflation, we consider the effective Lagrangian in the vacuum and discuss the unitarity problem.

By expanding the $\phi$ field around the vacuum with $\phi=\phi_0+{\bar\phi}$, we obtain the Lagrangian with $f(\phi)=\phi^n$,  $g(\phi)=\phi^m$ and $h(\phi)=\phi^k$ in eq.~(\ref{lag1}) as
\bea
\frac{{\cal L}}{\sqrt{-g}}= \frac{1}{2}\Big(1+\zeta(\phi_0+{\bar\phi})^k\Big) R -\frac{1}{2}\Big(1+\xi(\phi_0+{\bar\phi})^n\Big)(\partial_\mu{\bar\phi})^2 -\lambda \Big((\phi_0+{\bar\phi})^m-\phi^m_0\Big)^2.
\eea
Then, after the scalar perturbation is canonically normalized by ${\hat\phi}=\sqrt{1+\xi\phi^n_0}\,{\bar\phi}$,
the above Lagrangian is rewritten as
\bea
\frac{{\cal L}}{\sqrt{-g}}&=&\frac{1}{2}R+\frac{1}{2}\zeta\phi^k_0\left(1+\frac{{\hat\phi}}{\phi_0\sqrt{1+\xi\phi^n_0}}\right)^k R
\nonumber \\
&&-\frac{1}{2}\left(1+\frac{\xi\phi^n_0}{1+\xi \phi^n_0}\left[\left(1+\frac{{\hat\phi}}{\phi_0\sqrt{1+\xi\phi^n_0}}\right)^n-1 \right]\right)(\partial_\mu{\hat\phi})^2 \nonumber \\
&&-\lambda \phi^{2m}_0 \left[\left(1+\frac{{\hat\phi}}{\phi_0\sqrt{1+\xi\phi^n_0}}\right)^m-1\right]^2. \label{pertaction}
\eea
Therefore, for $\xi\phi^n_0\gtrsim 1$, the relevant higher order interactions for determining the unitarity scale are the following,
\be
\frac{\zeta}{(\xi \phi^n_0)^{1/2}}\,{\hat\phi}\, h^\nu_\nu \Box h^\mu_\mu\,\,(k=1),\quad \frac{\zeta}{(\xi \phi^n_0)^{k/2}}\,{\hat\phi}^k \Box h^\mu_\mu\,\,(k>1)  \label{hd1}
\ee
\be
 \frac{1}{(\xi \phi^{n+2}_0)^{n/2}}\,{\hat\phi}^n (\partial_\mu {\hat\phi})^2\,\,(n>0), \quad \frac{\lambda}{(\xi \phi^{n}_0)^m}{\hat\phi}^{2m}\,\,(m>2)  \label{hd2}
\ee
where the non-minimal coupling $\zeta$ is assumed to be of order one so the graviton kinetic term is already canonically normalized. 
From the first higher dimensional operator in eq.~(\ref{hd2}), we identify the unitarity scale $\Lambda_{UV}$ as follows,
\bea
\Lambda_{UV}=\frac{\sqrt{\xi} \phi^{1+n/2}_0}{M^{n/2}_P}.
\eea
Consequently, for $\Lambda_{UV}\sim M_P$,  we need the vacuum expectation value of the $\phi$ field to be 
\be
\phi_0\sim \frac{M_P}{\xi^{1/(n+2)}}.
\ee
For instance, requiring $\xi\gtrsim 10^{2(n+2)}$ for sub-Planckian field values of $\phi$ during inflation, we just need the inflaton VEV to be $\phi_0\lesssim 0.01 M_P$. We note that the non-minimal coupling does not lead to a lower unitarity scale than the Planck scale, as far as $\zeta \lesssim (\xi \phi^n)^{k/2}$. Moreover, the unitarity scale identified from the potential in eq.~(\ref{hd2}) is not smaller than the Planck scale, as far as $\lambda\lesssim (\xi \phi^n_0)^m$.  We note that for $\xi\phi^n_0\gtrsim 1$, the inflaton mass in the vacuum is the same value as during inflation, which is given by eq.~(\ref{inflatonmass}).

Now we address the validity of the classical approximation for inflation in view of the identified unitarity scale.  The bottom line is that once the classical Lagrangian with a large coupling in the kinetic term is given and the inflation VEV is large, the unrenormalized, non-canonical field is always sub-Planckian and the model is unitary and self-consistent at the quantum level up to the Planck scale. When higher dimensional operators in the scalar functions $h(\phi), f(\phi)$ and $g(\phi)$ are generated by new physics at the unitarity scale, that is, the Planck scale in our model, they could affect the classical inflationary dynamics.  But, since the unrenormalized, non-canonical $\phi$ field is sub-Planckian \footnote{The inflaton field value during inflation is greater than a naive cutoff $M_P/\xi^{1/n}$ expected for the vacuum with a small inflation VEV. But, as we discussed, the true cutoff depends on the inflaton VEV, becoming of order the Planck scale for a large inflaton VEV.},  those higher dimensional interactions for the $\phi$ field, if Planck-scale suppressed as in eq.~(\ref{hdo1}), are under control. 
Therefore, there is no extra degree of freedom needed to make the monomial chaotic inflation UV complete below the Planck-scale cutoff and our model predictions are insensitive to the Planck-scale suppressed interactions.
Thus, the model can be treated in an effective field theory consistently from the vacuum all the way to the inflationary era.

For general vacuum expectation values of the inflaton, the unitarity scale read off from the most dangerous operator, ${\hat\phi}^n (\partial {\hat \phi})^2$, in eq.~(\ref{pertaction}), becomes
\be
\Lambda_{UV}= \left(1+\frac{\xi \phi^n_0}{M^n_P}\right)^{\frac{1}{2}+\frac{1}{n}} \xi^{-1/n}M_P.
\ee
Therefore, if the vacuum expectation value of the inflaton field is negligible such that $\xi \phi^n_0\ll M^n_P$, the unitarity cutoff is saturated to $\Lambda_{UV}\approx\xi^{-1/n}M_P$, which is smaller than the Planck scale. On the other hand, during inflation, the inflaton $\phi$ runs over the field values, $\phi\gg \xi^{-1/n} M_P= \Lambda_{UV}$, which are beyond the unitarity cutoff. Therefore, in this case, higher order terms for $\phi$ suppressed by the Planck scale could not be ignored. For instance, for the Higgs inflation with running kinetic term \cite{runkinetic}, the vacuum expectation value of the Higgs field is small so that the non-canonical kinetic term with a large $\xi$ introduces a unitarity problem below the Planck scale, so the model is sensitive to higher order terms for the Higgs field suppressed by $\Lambda_{UV}=\xi^{-1/n}M_P$. Therefore, the Higgs inflation with running kinetic term is similar to the original Higgs inflation with non-minimal gravity coupling, where inflaton field values are beyond the unitarity cutoff  during inflation \cite{uvhiggs}.

\section{Conclusions}

We have proposed the general chaotic inflation models that share the predictions of the usual chaotic inflation with canonically-normalized inflaton field and are favored by the recent BICEP2 observation of a large tensor-to-scalar ratio. 
Focusing on the general quadratic inflation, we have shown that the non-canonical inflaton remains sub-Planckian during inflation while the unitarity scale identified in the vacuum is of order the Planck scale, after the non-canonical inflaton field obtains a large vacuum expectation value and it is canonically normalized.  Therefore, the higher order terms for the non-canonical inflaton field, that are suppressed by the Planck scale, are negligible. The very choice of particular terms in the classical Lagrangian should be justified for another reason, but we showed that there is a class of self-consistent attractor solutions in a general form leading to the sub-Planckian quadratic inflation with the Planck-scale cutoff. The results were generalized to monomial chaotic inflation models. 
Our result indicates that the well-known chaotic inflation models such as quadratic inflation can be UV complete  in the presence of the non-canonical kinetic term, due to the large vacuum expectation value of the inflaton.

\section*{Acknowledgments}

The work is supported in part by Basic Science Research Program through the National Research Foundation of Korea (NRF) funded by the Ministry of Education, Science and Technology (2013R1A1A2007919).  This research was supported by the Chung-Ang University Research Grants in 2014.

\end{document}